\documentclass[twocolumn,showpacs,preprintnumbers]{revtex4}
\usepackage{amsmath}

\input{tcilatex}

\begin{document}

\title{Monte-Carlo rejection as a tool for measuring the energy landscape
scaling of simple fluids}
\author{Gerardo G. Naumis}
\affiliation{Instituto de F\'{\i}sica, Universidad Nacional Aut\'{o}noma de M\'{e}xico
(UNAM), Apartado Postal 20-364, 01000, M\'{e}xico, Distrito Federal, Mexico.}
\date{\today }
\pacs{64.70.Pf, 61.20.-p, 68.18.Jk}

\begin{abstract}
A simple modification of the Monte-Carlo algortihm is proposed to explore
the topography and the scaling of the energy landscape. We apply this idea
to a simple hard-core fluid. The results for different packing fractions
show a power law scaling of the landscape boundary, with a characteristic
scale that separates the values of the scaling exponents. Finally, it is
shown how the topology determines the freezing point of the system due to
the increasing importance and complexity of the boundary.
\end{abstract}

\maketitle

\section{Introduction}

A liquid cooled to temperatures near its freezing point can be conduced to a
glassy state or to a crystal according to the speed of cooling \cite{Tabor}.
When the speed is high enough, the supercooled liquid undergoes a glass
transition to a state that is disordered, while a phase transition of the
type fluid-crystal is obtained if the system is kept in thermodynamical
equilibrium at all steps of the cooling path \cite{Debenedetti}. The
understanding of the many different aspects of glass transition still
remains as one of the most important problems in physics \cite{Anderson}, as
for example, the explanation of the non-exponential relaxation of
fluctuations \cite{Relaxation} or why not all materials form glasses \cite%
{Jackle}. Another very interesting property of glasses, related with the
glass forming tendency \cite{Tatsumisago}, is the behavior of the viscosity,
which is usually referred as fragility. Different approaches have been used
to understand glass transition : models like the Gibbs-DiMarzio \cite%
{Debenebook}, theories like the mode coupling, stochastic agglomeration \cite%
{Kerner1}\cite{Kerner2}\cite{Micoulaut} or the use of computer simulations 
\cite{Debenebook}. Another useful approach is the rigidity theory of
Phillips \cite{Phillips1} and Thorpe \cite{Thorpe0}, which relates the ratio
between available degrees of freedom and the number of constraints \cite%
{Thorpe1}. In previous works, we showed that even for simple systems like
hard-disks \cite{Huerta0} and colloids \cite{Huerta1}, it seems that
rigidity plays an important role even for the case of a simple phase
transition, since it is clear that in order to form a solid, the system must
develop certain rigidity. Some works on the relaxation properties of
colloids, seem to confirm these ideas \cite{Weitz}.

Parallel to all of these approaches, there is another formalism that has
been very useful to visualize and understand what happens during a glass or
usual phase transition . This formalism is the energy landscape approach 
\cite{Goldstein}\cite{Angell}, which many years ago was very successful in
the field of spin glasses \cite{Mezard}. The main idea behind this approach,
is that the landscape is a surface generated by the potential energy of the
system as a function of the molecular positions \cite{Debenedetti}. For a
system with $N$ molecules, the landscape is determined by the potential
energy function, $\Phi (\mathbf{r}_{1},...,\mathbf{r}_{N}),$ where $\mathbf{r%
}_{i}$ comprise all relevant coordinates, like position and orientation.
Since the kinetic energy ($K$) is a positive defined quantity, the system
evolves in such a way that $K=E-\Phi (\mathbf{r}_{1},...,\mathbf{r}_{N})\geq
0,$ where $E$ is the total energy. The topography of the landscape energy
surface determines the possible evolution of the system, and the contact
with thermodynamics is made by using statistical mechanics \cite{Debenedetti}%
. The great advantage of the landscape formalism is that it can be used even
without thermodynamical equilibrium. In such a case, ergodicity is broken
and the entropy is not a maximum anymore, as postulated in the usual
equilibrium statistical mechanics. The main question to be answered for this
case, is to figure out the fraction of allowed volume in phase space that is
visited by the system.

The usual picture of a phase or glass transition in such a language, is that
at high temperatures the system does not feel the topology of $\Phi (\mathbf{%
r}_{1},...,\mathbf{r}_{N})$ because the kinetic energy contribution
dominates. As the temperature is lowered, the system is unable to surmount
the highest energy barriers and therefore is forced to sample deep minima.
For a slow cooling is slow, the system has time to find a path to the most
stable state, an ordered crystal. It will be trapped in a metastable state,
the glass, \cite{Goldstein}\cite{Angell} if the cooling speed is high
enough. Many works that relates the statistics of an energy landscape with
the thermodynamical properties of the system have been made \cite%
{Debenedetti}\cite{Stillinger}\cite{Buchner}\cite{Shah}\cite{Shell}, and
even some phenomenological relations between rigidity and the energy
landscape have been obtained \cite{Naumis}. For a Lennard-Jones fluid, it
has been determined that the network of minima is a static scale-free
network \cite{Doye}.

However, up to now there are some important questions that remain, for
example, what is the topography of the landscape for a fragile liquid or how
to obtain the viscosity \cite{Angell}, diffusion \cite{Keyes} and rigidity
from the landscape \cite{Naumis}. Also, although is widely believed that the
landscape is fundamental to understand many features of a liquid, still is
not completely clear how to use the landscape topology to predict a phase
transition \cite{Fyodorov}. Another interesting question is: what is the
nature of the texture of the landscape? In other words, do the topography is
a fractal? What's the fractal dimension of the landscape? Although some of
these questions seem to have an academic interest only, is clear that the
relaxation properties of a very complex fractal landscape are different from
a smooth landscape \cite{Keyes2}, where the system can easily explore the
phase space. In this article, we will explore some of these questions by
looking at the scaling of the landscape.

The layout of this work is the following: in section II we discuss how to
use a modified Monte-Carlo method to study the scaling. In section III we
apply the method to a simple fluid of hard disks. Finally, in section IV we
give the conclusions.

\section{Monte-Carlo rejection and scaling of the landscape}

In this section we will develop a method to relate the Monte-Carlo rejection
ratio and the scaling of the landscape. Before going into the details, it
would be useful to explain some others approaches to obtain information
about the landscape topology. To simplify ideas, in this article we will use
as a model system, $N$ hard disks or spheres of diameter $\sigma $ in a
given area ($S$) or volume ($V$). In such hard-core particle system, the
energy landscape is formed by walls of infinite height that divide the
allowed and forbidden regions of the configurational phase space. If $%
\mathbf{r}_{i}$ is the position of a disk or sphere $i$, the allowed region
of the landscape is the set of points where, 
\begin{equation}
\left\Vert \mathbf{r}_{i}-\mathbf{r}_{j}\right\Vert \geq \sigma ,
\label{rij}
\end{equation}%
for all possible pairs $i$ and $j.$ The number of such pairs ($R_{NH}$) is
the number of combinations of $N$ objects taken in pairs: $C_{2}^{N}=N(N-1)/2
$. Each of these $C_{2}^{N}$ equations is a non-holonomic restriction to the
system. A state $\mathbf{P}$ in phase space is in the boundary of the
landscape, if at least one of the inequalities (\ref{rij}) is transformed
into an holonomic restriction, 
\begin{equation}
\left\Vert \mathbf{r}_{l}-\mathbf{r}_{m}\right\Vert =\sigma ,  \label{eqrij}
\end{equation}%
for a pairs of disks that we denote by $l$ and $m$. For each equation of
this type that is satisfied, two disks are in contact. For a given packing
fraction ($\phi $), the number of such equations ($R_{H}$) is just, 
\begin{equation}
R_{H}=\frac{\left\langle Z(\phi )\right\rangle }{2}N,
\end{equation}%
where $\left\langle Z(\phi )\right\rangle $ is the average coordination per
particle in a given packing. We remark that this equation allows a straight
forward manner to connect the energy landscape formalism with the rigidity
theory, where the most important parameter \cite{Phillips1} is $\left\langle
Z(\phi )\right\rangle $. This approach also provides a way to construct
inherent structures and the boundary of the landscape just by considering a
non-linear optimization problem. To get a packing, we can define an
objective function as, 
\begin{equation}
F(\mathbf{P})=\dsum\limits_{j=1}^{N}\left\Vert \mathbf{r}_{i}\right\Vert ,
\label{F(p)}
\end{equation}%
with $R_{H}$ constraints $\left\Vert \mathbf{r}_{l}-\mathbf{r}%
_{m}\right\Vert =\sigma .$ This objective function is defined in such a way
that the particles are packed in a tight way with respect to the origin.
Surprisingly, this criteria is similar to the center of mass minimization
that has been observed very recently in experiments with colloids \cite%
{Science}\cite{Manoharan}. In principle, this problem can be solved using
nonlinear programming \cite{Saaty}, and there is some similar work done into
this line of research \cite{Corty}. Here we will not follow this path.
Instead, we investigate how the landscape boundary looks at different scales
in the configurational part of the phase space. The most simple way to do
this, consists in applying a box-counting algorithm \cite{Falconer} once the
boundary points are determined. In this box counting algorithm, a grid made
from cubes of linear size $\delta $ is applied to the configurational phase
space. Then, the number of boxes that contains a boundary state are counted.
The counting is repeated at different lengths $\delta $. This ideal
situation has the problem that we need to generate all the boundary states,
and due to the size of the phase space, this task is almost impossible to
do. A simpler approach is to take advantage of the Monte-Carlo importance
sampling to obtain information about the boundary. \bigskip 

In a usual Metropolis Monte-Carlo \cite{Ciccotti}\cite{Binder}, when a
system is in a microstate $\mathbf{P}_{0},$ a movement to a new microstate $%
\mathbf{P}$ is allowed if the difference in energies $\Delta E=E(\mathbf{P}%
)-E(\mathbf{P}_{0})$ is negative (where $E(\mathbf{P})$ is the energy of the
state $\mathbf{P}$). If $\Delta E>0,$ a random number is compared with $%
\Delta E.$ In the case of hard-core systems, a rejection occurs when a new
proposed configuration is not allowed by the restrictions \cite{Binder}, 
\textit{i.e.}, when there is an overlap between disks or spheres. When the
new proposed point $\mathbf{P}$ is rejected, one can be sure that the
boundary of the landscape has been crossed, \textit{i.e.}, the boundary is
between states $\mathbf{P}_{0}$ and $\mathbf{P.}$ Thus, the information
about the boundary can be extracted from the acceptance ratio of the
Monte-Carlo, although two modifications are needed. In the usual
Monte-Carlo, the trial movement distance between two states is a continuous
random variable \cite{Allen}. This feature is not convenient because it does
not provide an approximate location for the boundary. A second fact to take
into account, is that the probability of hitting the point $\mathbf{P}$ not
only depends on the size of the boundary, but also in the transient
probability of the process \cite{Jain} $\mathbf{P}_{0}\rightarrow \mathbf{P}$%
. To solve these problems, let us introduce a regular grid in the
configurational part of the phase space. If the simulation is performed in a
box of linear length $L,$ there are $M=(L/\delta )^{DN}$ points in the grid,
where $D$ is the dimensionality of the system. In such a grid, a random walk
in phase space is performed by choosing at random a disk $\mathbf{r}_{i}$
and changing one components at random to $\pm \delta .$ If $\mathbf{P}$ is
written in generalized coordinates $q_{j}$, the trial movement is written
as, 
\begin{equation}
(q_{1},q_{2},q_{3},...,q_{DN})\rightarrow (q_{1},q_{2},q_{3},...,q_{r}\pm
\delta ,q_{DN}),  \label{q}
\end{equation}%
for the $r$ coordinate chosen at random with an uniform distribution.

The introduction of a random walk is convenient because: 1) the step size
can be varied to look at the scaling and 2) when a movement is rejected, the
state can be considered as a boundary point, since it is connected to the
interior of the forbidden part of the phase space. In spite of this, the
introduction of a random walk has the inconvenient of not being able to
sample the phase space with equal probabilities,\textit{\ i.e.}, it is not
completely ergodic. In fact, even for the usual Monte-Carlo there is some
lack of ergodicity due to the existence of an finite-size underlying grid.
The only difference between the simple random walk and the usual Monte-Carlo
with jumps of variable size is the higher interconnectivity of the grid in
the former case, which mitigates but not solves completely the effects of
the biased sampling. The main effect of this problem upon our calculations,
it is that the size of the boundary will be underestimated, and that some
parts of the landscape are not going to be visited. Thus, the method works
better before a transition, and in fact only provides a bound for the
scaling exponents. To improve upon this method, some other algorithms can be
used, as for example collision prediction \cite{Lubachevsky}.

Now let $p_{k}(\delta )$ be the probability of state $k$ in phase space to
be occupied by the system when a grid of scale $\delta $ is chosen. The
random walk process can be viewed as a Markov chain \cite{Jain}, where the
probabilities of visiting each microstate are contained in a vector . The
probabilities at each step are transformed according to an stochastic matrix
that contains as elements the probability of transition among states, 
\begin{equation*}
\left( 
\begin{array}{c}
p_{1}^{\prime }(\delta ) \\ 
p_{2}^{\prime }(\delta ) \\ 
... \\ 
p_{M}^{\prime }(\delta )%
\end{array}%
\right) =\left( 
\begin{array}{cccc}
S_{11} & S_{12} & ... & S_{1M} \\ 
S_{21} & S_{22} & ... & S_{2M-1} \\ 
... & ... & ... & ... \\ 
S_{M1} & ... & ... & S_{MM}%
\end{array}%
\right) \left( 
\begin{array}{c}
p_{1}(\delta ) \\ 
p_{2}(\delta ) \\ 
... \\ 
p_{M}(\delta )%
\end{array}%
\right) ,
\end{equation*}%
where the rows of the matrix are normalized to $1,$ and $p_{k}^{\prime
}(\delta )$ are the probabilities after a step is made. In a hard-core
system, jumps only occur between allowed grid points that are first
neighbours. An element $S_{rt}$ of this matrix is zero when one of the
states $r$ or $t$ is in the forbidden part of the landscape. $S_{rt}$ is
also zero if $r$ or $t$ are not first neighbours. The only elements
different from zero are $S_{rt}=1/z_{t}(\delta )$, if $r$ and $t$ are
allowed neighbours, where $z_{t}(\delta )$ is the coordination in phase
space of site $t$ (i.e., the number of allowed first neighbours of $t$) for
a given scale $\delta $. Points at the boundary of the landscape are the
ones where $1/z_{r}(\delta )<2DN$ since they are connected to points inside
the forbidden part of the phase-space. A matrix of this type has at least
one eigenvector with eigenvalue one, while the others have norms equal or
less than one \cite{Kerner2}. Thus, after successive applications of the
matrix, the stable configuration is given by the eigenvalue with norm one,
from where it follows that the final probabilities satisfy \cite{Kerner2}, 
\begin{equation*}
\left( 
\begin{array}{c}
p_{1}(\delta ) \\ 
p_{2}(\delta ) \\ 
... \\ 
p_{M}(\delta )%
\end{array}%
\right) =\left( 
\begin{array}{cccc}
S_{11} & S_{12} & ... & S_{1M} \\ 
S_{21} & S_{22} & ... & S_{2M-1} \\ 
... & ... & ... & ... \\ 
S_{M1} & ... & ... & S_{MM}%
\end{array}%
\right) \left( 
\begin{array}{c}
p_{1}(\delta ) \\ 
p_{2}(\delta ) \\ 
... \\ 
p_{M}(\delta )%
\end{array}%
\right)
\end{equation*}%
This matrix is similar to the Hamiltonian of a binary alloy in an hypercubic
lattice, where the two self-energies are very different \cite{Binary} (split
band regimen). It is easy to prove that the final equilibrium vector
coincides with the bonding state (which corresponds with the maximal
wave-length of the solution, and nearly zero phase difference between sites)
of the binary alloy.

In a Monte-Carlo step, the probability of having a rejection is given by the
probability of jumping into a boundary point ($p_{k}(\delta )$), multiplied
by the probability of jumping form a state $k$ into a state $t$ in the
forbidden region of the landscape, which is given by the elements of the
stochastic matrix. Thus, the probability of landing in a forbidden state $t$
is, 
\begin{equation}
p_{t}(\delta )=\left( \frac{2DN-z_{k}(\delta )}{2DN}\right) p_{k}(\delta ).
\end{equation}
If $\mathcal{L}_{B}(\delta )$ denotes the set of all boundary points, the
total probability of having rejections at a scale $\delta $ (we denote this
probability by $p^{R}(\delta )$) is obtained by summing over all boundary
states $k$, 
\begin{equation}
p^{R}(\delta )\equiv \dsum\limits_{k\in \mathcal{L}_{B}(\delta )}\left( 1-%
\frac{z_{k}(\delta )}{2DN}\right) p_{k}(\delta ).  \label{M}
\end{equation}%
Now we write $z_{k}(\delta )$ as an average plus a fluctuation part, $%
z_{k}(\delta )=<z_{k}(\delta )>+\Delta z_{k}(\delta )$, where 
\begin{equation}
\left\langle z(\delta )\right\rangle =\frac{1}{M_{B}(\delta )}%
\dsum\limits_{k\in \mathcal{L}_{B}(\delta )}z_{k}(\delta ),
\end{equation}%
and $M_{B}(\delta )$ is simply the number of boundary points at a scale $%
\delta .$ The same procedure can be made for $p_{k}(\delta )$, giving $%
p_{k}(\delta )=\left\langle p_{B}(\delta )\right\rangle +\Delta p_{k}(\delta
),$where $\left\langle p_{B}(\delta )\right\rangle $ is defined as, 
\begin{equation}
\left\langle p_{B}(\delta )\right\rangle =\frac{1}{M_{B}(\delta )}%
\dsum\limits_{k\in \mathcal{L}_{B}(\delta )}p_{k}(\delta ).
\end{equation}%
Using these definitions, and that the sum of the fluctuations is zero, eq.(%
\ref{M}) is rewritten as, 
\begin{eqnarray*}
p^{R}(\delta ) &=&\dsum\limits_{k\in \mathcal{L}_{B}(\delta )}\left( 1-\frac{%
\left\langle z(\delta )\right\rangle }{2DN}\right) \left\langle p_{B}(\delta
)\right\rangle \\
&&+\dsum\limits_{k\in \mathcal{L}_{B}(\delta )}\left( \frac{\Delta
z_{k}(\delta )\Delta p_{k}(\delta )}{2DN}\right) .
\end{eqnarray*}%
The term $\dsum \Delta z_{k}(\delta )\Delta p_{k}(\delta )$ is a measure of
the correlation between state and coordination fluctuations. Since the
eigenvector with eigenvalue one corresponds to a bonding state in a binary
alloy, using a variational procedure with a trial function or analyzing the
spectral moments \cite{Binary}, it can be proved that $\Delta p_{k}(\delta
)\approx \Delta z_{k}(\delta )/2DN.$ This term gives a correction of order, 
\begin{equation}
\frac{1}{2DN}\dsum\limits_{k\in \mathcal{L}_{B}(\delta )}\Delta z_{k}(\delta
)\Delta p_{k}(\delta )\approx \left( \frac{\widehat{\sigma }(\delta )}{2DN}%
\right) ^{2},
\end{equation}%
where $\widehat{\sigma }(\delta )$ is the standard deviation of the phase
space coordination distribution $z_{k}(\delta )$. Thus, it follows that, 
\begin{equation}
M_{B}(\delta )<p_{B}(\delta )>=\left( \frac{p^{R}(\delta )-\left( \frac{%
\widehat{\sigma }(\delta )}{2DN}\right) ^{2}}{\left( 1-\frac{\left\langle
z(\delta )\right\rangle }{2DN}\right) }\right) ,  \label{Mb}
\end{equation}%
We notice that $M_{B}(\delta )<p_{B}(\delta )>$ is a bound for the size of
the whole boundary of the landscape. For example, when the sampling is
uniform, $M_{B}(\delta )<p_{B}(\delta )>=M_{B}(\delta )/M(\delta ),$ since $%
<p_{B}(\delta )>=1/M(\delta )$. The number of grid points $M(\delta )$
scales as $\delta ^{-DN}$, and if the boundary has a scaling of the type $%
M_{B}(\delta )\sim \delta ^{-D_{B}N}$, then we expect a scaling of eq.(\ref%
{Mb}) as, 
\begin{equation}
\frac{M_{B}(\delta )}{M(\delta )}\propto (\delta /\sigma )^{D_{f}},
\end{equation}%
where $D_{f\text{ }}=(D-D_{B})N$ is an effective fractal dimension due to
the different scalings of the grid and the landscape. In general, states at
the boundary are less visited, thus we get the following inequality, 
\begin{equation}
M_{B}(\delta )\left\langle p_{B}(\delta )\right\rangle \leq \frac{%
M_{B}(\delta )}{M(\delta )}.
\end{equation}%
As a result, we get a bound for this scaling exponent, 
\begin{equation}
D_{f}\leq \ln \left( \frac{p^{R}(\delta )-\left( \frac{\widehat{\sigma }%
(\delta )}{2DN}\right) ^{2}}{\left( 1-\frac{\left\langle z(\delta
)\right\rangle }{2DN}\right) }\right) /\ln (\delta /\sigma ).
\end{equation}%
The evaluation of this bound can be easily implemented inside a Monte-Carlo
simulation; it only requires the rejection probability $p^{R}(\delta )$, the
average coordination $<z(\delta )>,$ and the fluctuations $\widehat{\sigma }%
(\delta )$. To do this, first we divide the phase space with a grid of
spacing $\delta $. Then we perform the Monte-Carlo simulation, but if there
is a rejection during a trial step, this means that the initial state is in
the boundary of the landscape. To look at the coordination in phase space of
this state, a movement in each of the $DN$ coordinates of the gird is
performed, as in eq.(\ref{q}), but for each coordinate $q_{r}$ taken in
sequence from $r=1$ to $r=DN$. For each coordinate movement, the \ new state
is tested in order to determine if its an allowed or forbidden state. After
this cycle in the coordinates, the number of accepted states is the
coordination number $z_{k}(\delta )$. The process continues until a new
rejection appears, and at the end of the simulation, the average and the
standard deviation of the distribution of $z_{k}(\delta )$ are obtained. The
same procedure is repeated for different scales $\delta $.

Figure 1 illustrates the procedure for a very simple system. Consider two
disks in a box of length $L$ and width $\sigma $. In such a case, the
movement is one dimensional. If $x_{1}$ and $x_{2}$ are the coordinates of
each disk, the configurational part of the phase space can be represented as
a plane. The shape of the landscape is determined by the condition of
non-overlap $\left\vert x_{1}-x_{2}\right\vert \geq \sigma $ and the walls
of the box. The allowed phase space is made by two triangles, as shown in
fig. 1. Notice that ergodicity is always broken since the two triangular
regions are not connected. In fig. 1, the grid is indicated as dotted lines;
the points at the boundary (open circles in fig. 1) are those connected to
grid points that are outside the triangular regions (closed circles).

\FRAME{ftbpFU}{2.1404in}{2.5391in}{0pt}{\Qcb{Two disks in a rectangular box
of \ length $L$ and width $\protect\sigma $. Below the box, the
corresponding configurational part of the phase space is shown. The allowed
part of the landscape is the area indicated with the grey shadow. A grid of
scale $\protect\delta $ is indicated by dotted lines. Boundary points are
indicated by open circles. Closed circles are states in the forbidden part
of the phase-space. Notice that in this problem, ergodicity is always
broken, since the allowed parts of the landscape are not connected.}}{}{%
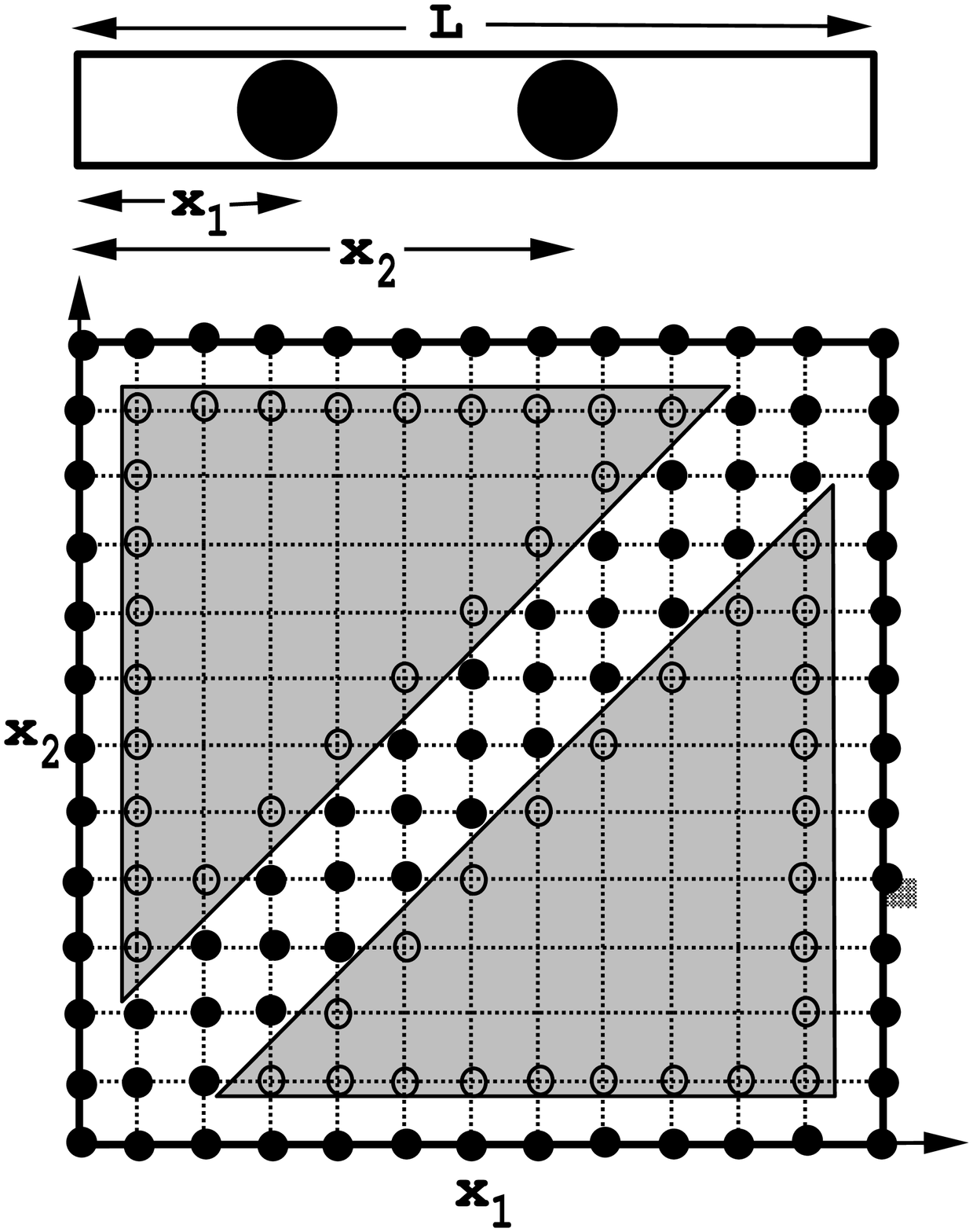}{\special{language "Scientific Word";type "GRAPHIC";display
"USEDEF";valid_file "F";width 2.1404in;height 2.5391in;depth
0pt;original-width 7.2004in;original-height 9.3478in;cropleft "0";croptop
"1";cropright "1";cropbottom "0";filename 'figure1.eps';file-properties
"XNPEU";}}

In this simple system, it is very instructive to compare the rejection ratio
of the Monte-Carlo with the topology of the phase space, since the
theoretical value for $M_{B}(\delta )/M(\delta )$ for a given packing
fraction $\phi =2/L$ is easy to find. The value of $M_{B}(\delta )/M(\delta )
$ in this case is given by the ratio between perimeter-area of the triangle
as a function of the scale, 
\begin{equation*}
\left[ \frac{M_{B}(\delta )}{M(\delta )}\right] _{L}=\left( \frac{2(2+\sqrt{2%
})}{L(1-2\sigma /L)}\right) \delta ,
\end{equation*}%
where the subscript is used to indicate that the result depends on the
corresponding length of the box. This result can be related with the
probability of rejection of the Monte-Carlo as follows. If we suppose an
uniform sampling, the probability of hitting a boundary point is given by
the perimeter-area ratio of the triangles. The average coordination of the
boundary points can be obtained from a direct inspection of the drawings
with different grids, that gives $\left\langle z(\delta )\right\rangle
\simeq \sqrt{2}+(3/2)$ for $\sigma \ll L$. We can neglect the term $\widehat{%
\sigma }(\delta )/4$, since it is very small to be considered (this
approximation was confirmed afterwards by the coordination statistics
obtained from the Monte-Carlo simulation). Using eq. (\ref{Mb}), the
predicted rejection probability is, 
\begin{equation*}
p^{R}(\delta )=m(L)\delta ,
\end{equation*}%
where $m(L)$ is defined as,%
\begin{equation*}
m(L)\equiv (\frac{5}{8}-\frac{1}{2\sqrt{2}})\left[ M_{B}(\delta )/M(\delta )%
\right] _{L}.
\end{equation*}%
The rejection is thus expected to be proportional to $\delta ,$ as confirmed
in Figure 2 by the numerical simulations using a Monte-Carlo, where the
rejections are plotted against $\delta $ for different $L$.\FRAME{ftbpFU}{%
3.9479in}{3.3754in}{0pt}{\Qcb{Rejection ratio as a function of the scale $%
\protect\delta $ (measured in units of $\protect\sigma $) for $L=50\protect%
\sigma $ (squares)$,30\protect\sigma $ (circles)$,15\protect\sigma $
(diamonds)$,10\protect\sigma $ (triangles up) and $4\protect\sigma $
(triangles down).}}{}{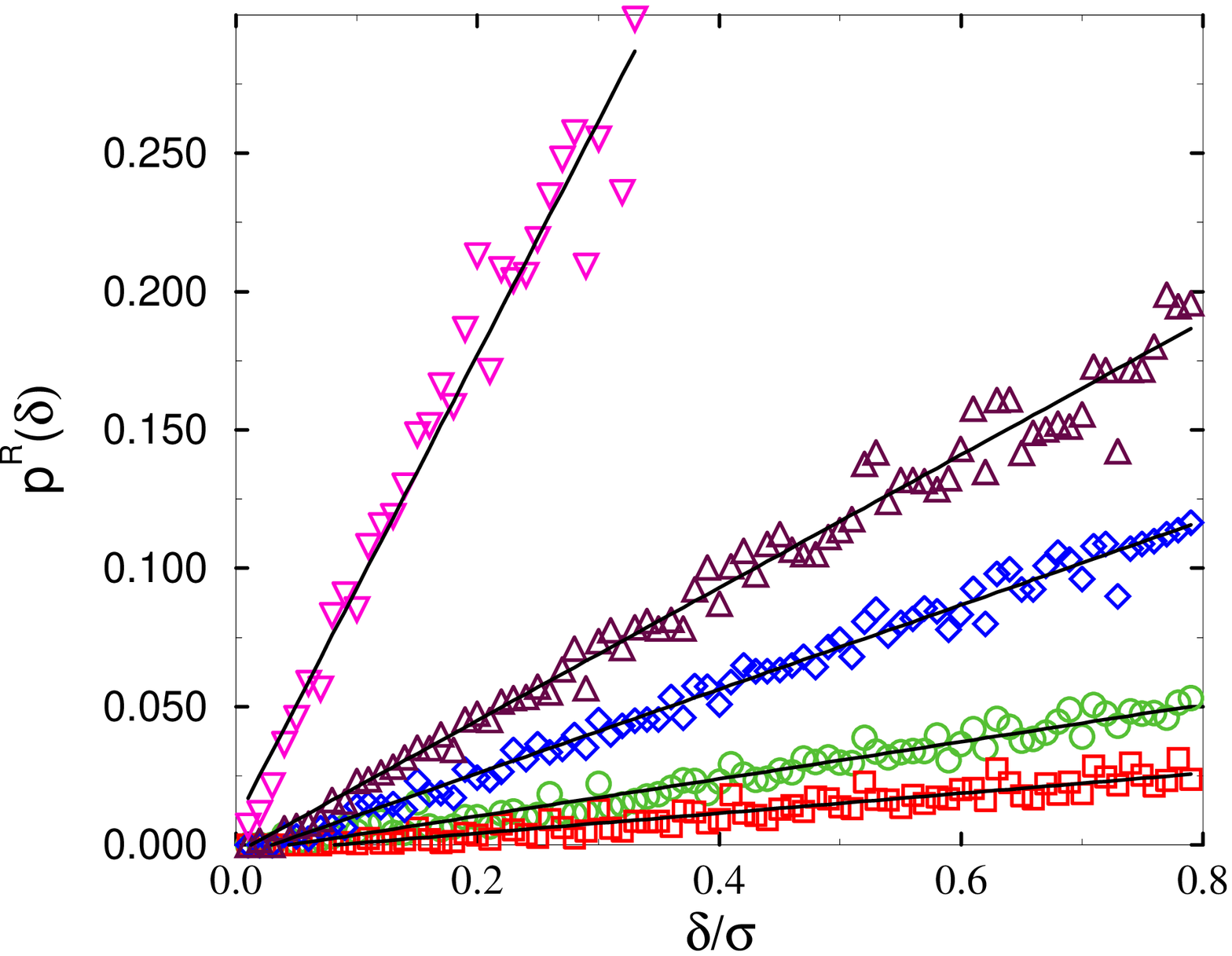}{\special{language "Scientific Word";type
"GRAPHIC";maintain-aspect-ratio TRUE;display "USEDEF";valid_file "F";width
3.9479in;height 3.3754in;depth 0pt;original-width 7.8369in;original-height
6.6919in;cropleft "0";croptop "1";cropright "1";cropbottom "0";filename
'figure2.eps';file-properties "XNPEU";}}\bigskip 

Using a least-square fitting, the slopes for each of the lines is shown in
figure 3 using a log-log plot. The solid curve is the theoretical value of $%
m(L)$ and the squares are the results of the simulation using the
Monte-Carlo simulation. These results are in very good agreement with the
theoretical values, specially for $\sigma \ll L$, where $m(L)$ is well
approximated by, 
\begin{equation*}
m(L)\approx \frac{8(2+\sqrt{2})}{L}(1+2\sigma /L),
\end{equation*}%
so $m(L)\sim L^{-1}$, and $D_{f}=1$ as expected for a smooth surface. In the
region $\sigma \approx L$, the difference between both results is due to the
fact that the average coordination number is not anymore $\sqrt{2}+(3/2)$,
and a correction is needed in the analytical formula. Also, in this region
the sampling is far from uniform, since the grid is very small compared with
the size of the boundary.

\FRAME{ftbpFU}{3.4186in}{3.5691in}{0pt}{\Qcb{Slopes of the fitting lines
that appear in fig. 2 as a function of $L.$ The solid line is the function $%
m(L).$}}{}{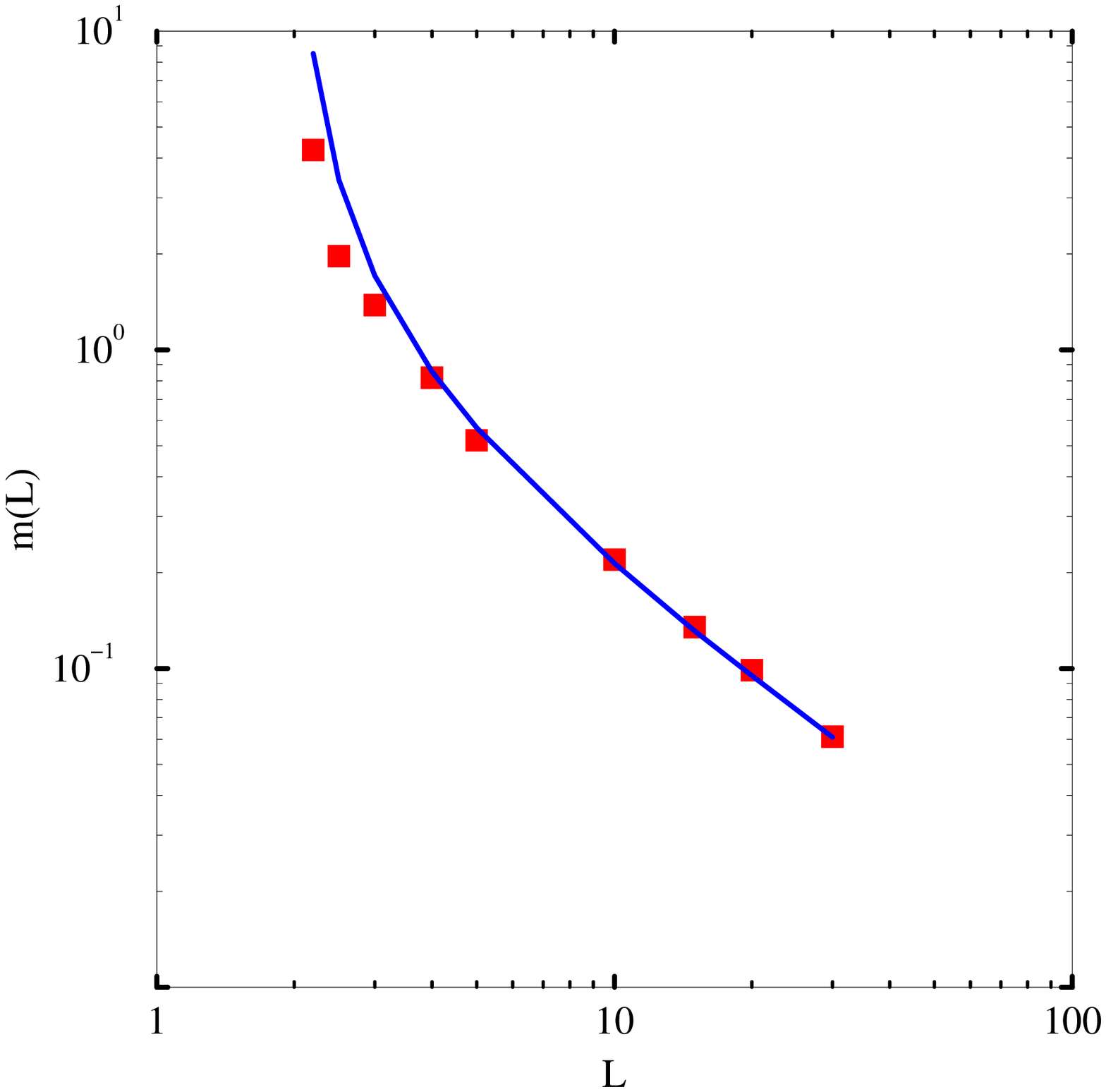}{\special{language "Scientific Word";type
"GRAPHIC";maintain-aspect-ratio TRUE;display "USEDEF";valid_file "F";width
3.4186in;height 3.5691in;depth 0pt;original-width 7.8343in;original-height
8.1829in;cropleft "0";croptop "1";cropright "1";cropbottom "0";filename
'figure3.eps';file-properties "XNPEU";}}

\section{Scaling in a simple liquid}

In this section, we show the results obtained using the method proposed in
the previous section for a two-dimensional system composed of hard-disks at
different packing fractions $\phi =N\pi \sigma ^{2}/4S,$ where $N=100$
particles. The thermodynamics of this system has been investigated by many
different groups during the last 50 years \cite{Metropolis}\cite{Alder}\cite%
{Hoover}. Despite the simplicity of the model, the nature of the phase
transition from solid to fluid is still debated \cite{Sengupta}, as well as
the nature of local order \cite{Stillinger2} and its relation with some
peaks in the radial distribution function \cite{Truskett}. Here we will only
investigate the landscape scaling. Figure 4 shows a log-log plot of $%
M_{B}(\delta )<p_{B}(\delta )>$ as a function of $\delta $ for different
packing fractions, as indicated in the figure caption. \FRAME{ftbpFU}{%
3.0666in}{2.5296in}{0pt}{\Qcb{Parameter $M_{B}(\protect\delta )\left\langle
p_{B}(\protect\delta )\right\rangle $ as a function of the scale $\protect%
\delta $, for different packing fractions. From top to bottom, $\protect\phi %
=0.74$ (squares)$,\protect\phi =0.71$ (x)$,\protect\phi =0.39$ (triangles), $%
\protect\phi =0.23$ (stars), $\protect\phi =0.12$ (diamonds), $\protect\phi %
=0.08$ (squares) and $\protect\phi =0.04$ (filled circles). The lines were
obtained using a power law fit.}}{}{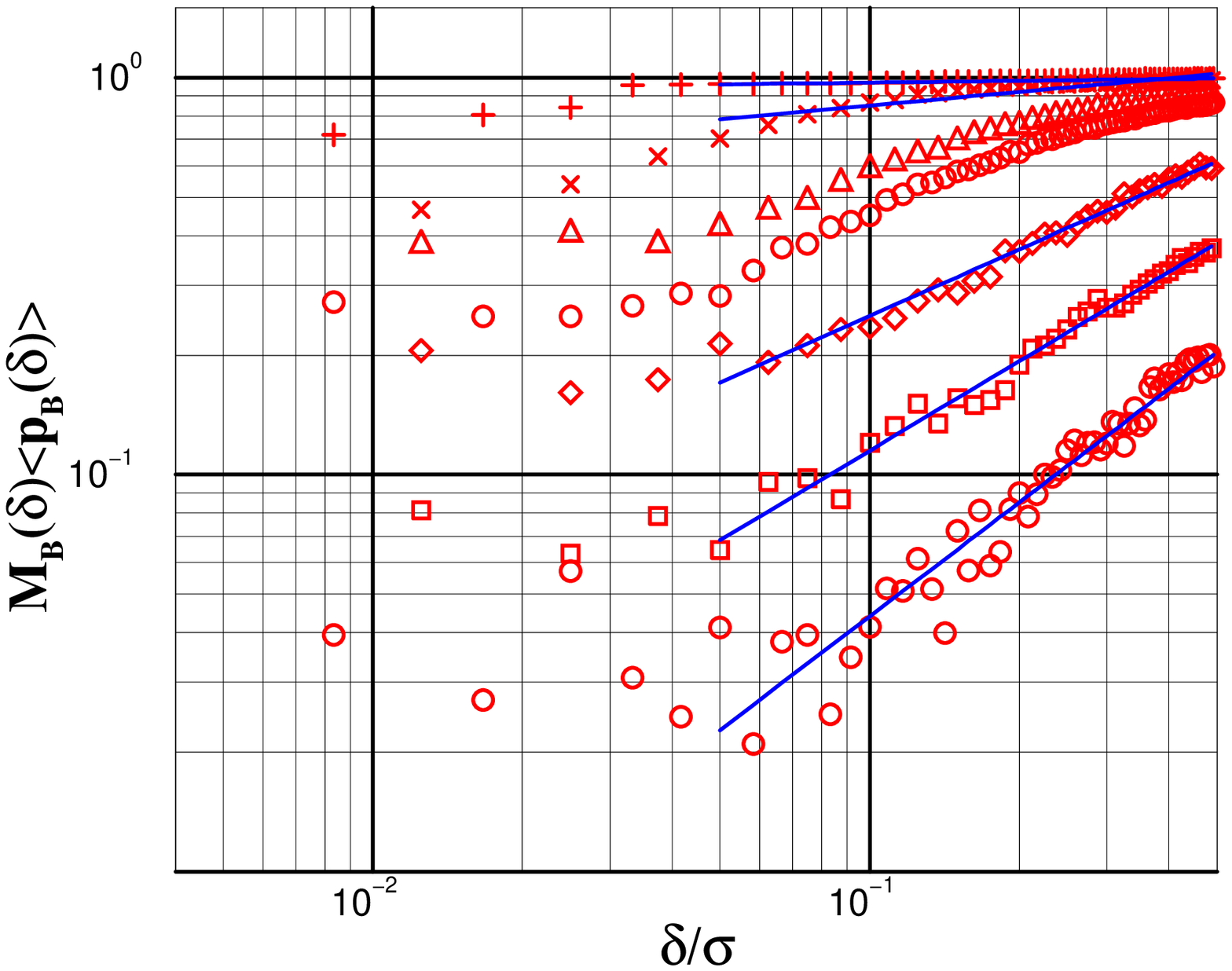}{\special{language
"Scientific Word";type "GRAPHIC";maintain-aspect-ratio TRUE;display
"USEDEF";valid_file "F";width 3.0666in;height 2.5296in;depth
0pt;original-width 7.8369in;original-height 6.4558in;cropleft "0";croptop
"1";cropright "1";cropbottom "0";filename 'figure4.eps';file-properties
"XNPEU";}}

Figure 5 is a similar plot, but only for packing fractions near the freezing
point (denoted by $\phi _{0}$). Both plots give evidence that there is a
power law scaling of $M_{B}(\delta )<p_{B}(\delta )>$ with $\delta $. This
power law behavior is clear near the freezing point or at low densities,
where fits of the type $\delta ^{D_{f}}$ are shown for the different sets of
data (only fits with correlation coefficients bigger than $99\%$ are shown).
Notice that all the fits have a cut-off at $\delta =0.05\sigma $, since
there is a cross-over in the power law behavior, \textit{i.e.}, for a given
packing fraction, two regions with different scaling exponents are observed,
as seen in Figure 5, where a drop of $M_{B}(\delta )<p_{B}(\delta )>$ is
observed around $\delta =0.05\sigma $. For low packing fractions, the
exponents for $\delta <0.05\sigma $ tend to be smaller than in the region $%
\delta >0.05\sigma $. The fact that two exponents are observed means that
below a certain length-scale, the landscape has a different structure. For
all the different packing fractions, this behavior is nearly similar. We can
speculate that this change of regimen for the scaling at a length-scale is
related with the different processes of relaxation that have been observed
in diverse simulations\FRAME{ftbpFU}{2.9689in}{2.5382in}{0pt}{\Qcb{
Parameter $M_{B}(\protect\delta )\left\langle p_{B}(\protect\delta %
)\right\rangle $ close to freezing as a function of $\protect\delta $, at $%
\protect\phi =0.74$ (squares) and $\protect\phi =0.63$ (circles).}}{}{%
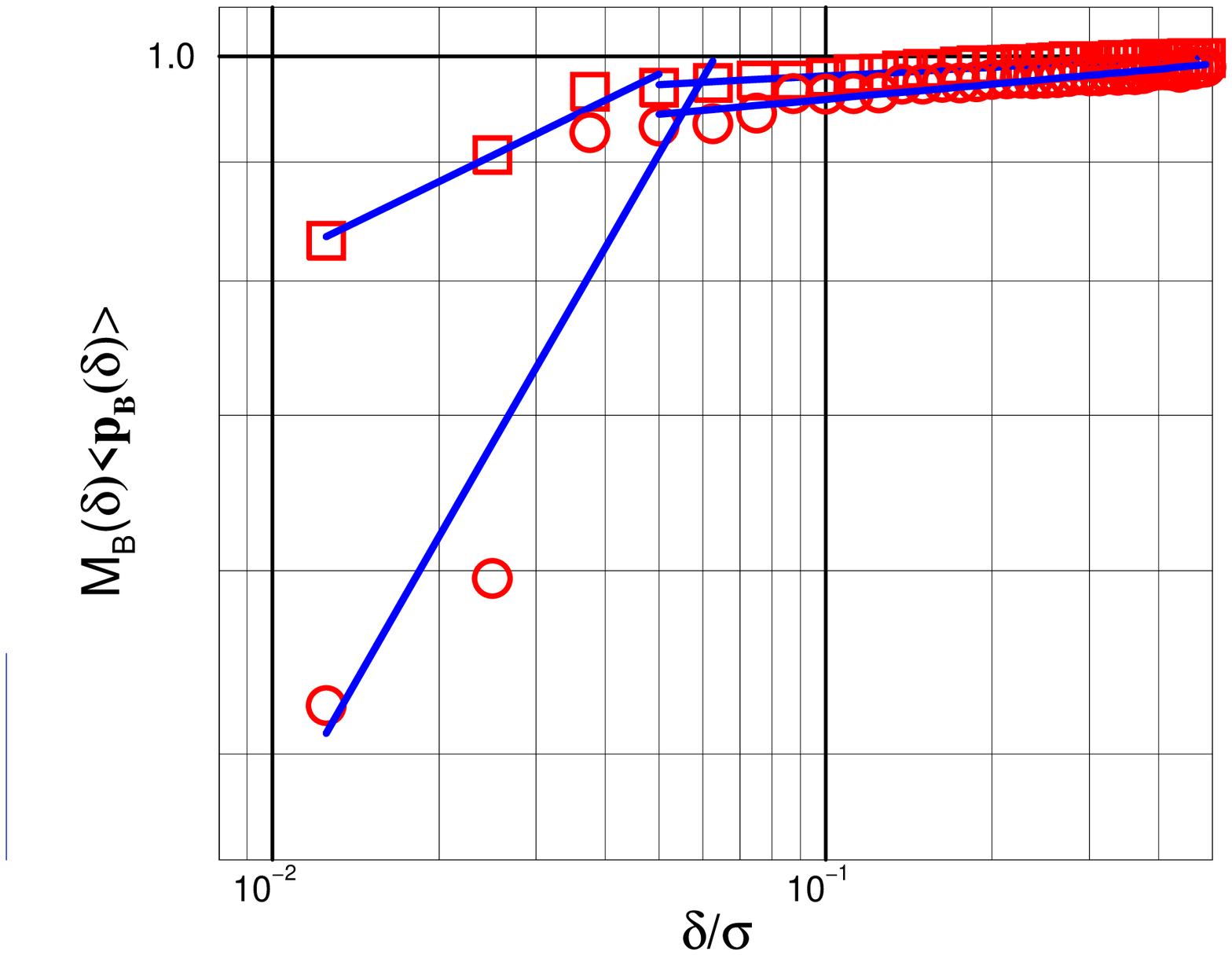}{\special{language "Scientific Word";type
"GRAPHIC";maintain-aspect-ratio TRUE;display "USEDEF";valid_file "F";width
2.9689in;height 2.5382in;depth 0pt;original-width 7.8369in;original-height
6.6919in;cropleft "0";croptop "1";cropright "1";cropbottom "0";filename
'figure5.eps';file-properties "XNPEU";}} \cite{Keyes2}\cite{Speedy}\cite%
{Erpenbeck} and experiments \cite{Weitz}, since although a Monte-Carlo
simulation does not provide the real dynamic of the system, is clear that a
big length scale $\delta $ in phase space correspond to long times in the
evolution of the system, as also expected from the Adam-Gibbs relation
between relaxation times and configurational entropy \cite{Debenebook}.
However, this speculation needs to be investigated in more detail.

We also notice that for packing fractions $0.2<\phi /\phi _{0}<0.6$, it
seems that using one single scaling exponent is not enough to fit the data,
which is an indicative of a multifractal structure, although if we restrict
the fitting for $\delta >0.2$, again a good power law fit is obtained.

In figure 6 we plot the scaling exponents obtained from the data of fig. 4
and 5 as a function of the packing fraction, for the regions ($0.2\sigma
<\delta $) where a clear scaling is obtained for all the graphs . As shown
in the figure, as the packing fraction reaches the freezing point, $D_{f}$ \
goes to zero, and the landscape boundary scales nearly as the volume in
phase space. This means that near the freezing point, the topology of the
landscape restricts in a severe manner the available phase space. Thus,
figure 6 provides clear evidence of how the topology of the landscape is
responsible for the phase transition that occurs at the freezing point, and
reinforces the speculation about relaxation times, since it has been
observed in experiments with colloids that freezing occurs when long time
relaxation is not longer available \cite{Weitz}.\FRAME{ftbpFU}{3.3676in}{%
2.6576in}{0pt}{\Qcb{Exponent $D_{f}$ as a function of the ratio between the
packing fraction and the packing fraction at freezing ($\protect\phi _{0}$).
The size of the squares is proportional to the maximal error, and the line
is a visual guide to the eye.}}{}{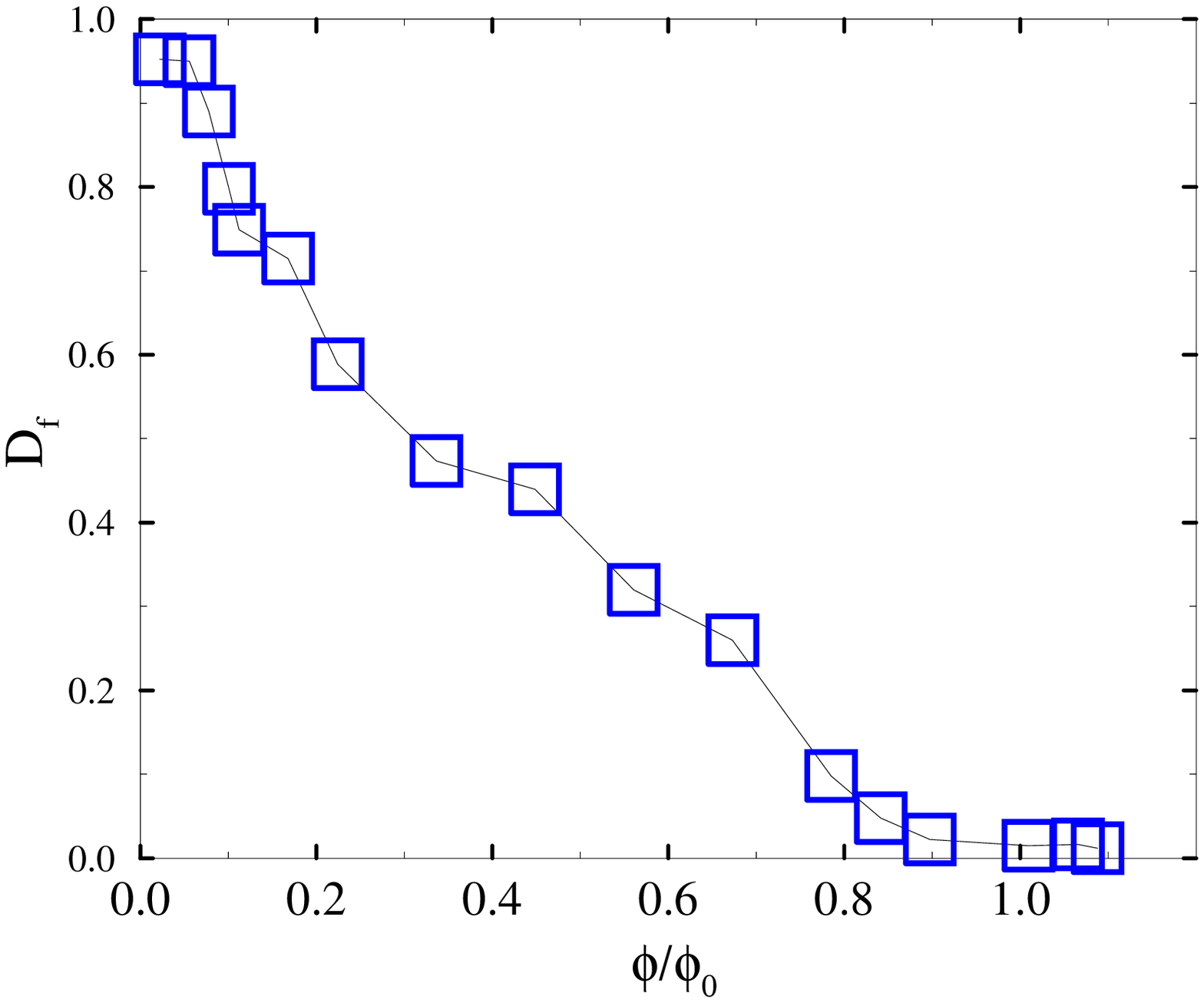}{\special{language "Scientific
Word";type "GRAPHIC";maintain-aspect-ratio TRUE;display "USEDEF";valid_file
"F";width 3.3676in;height 2.6576in;depth 0pt;original-width
7.8369in;original-height 6.1756in;cropleft "0";croptop "1";cropright
"1";cropbottom "0";filename 'figure6.eps';file-properties "XNPEU";}}

\section{Conclusions}

In this article, we have discussed some aspects of how to characterize the
structure and texture of the energy landscape in simple fluids. As a result,
we showed a method to investigate the boundary of the landscape that uses
the Monte-Carlo rejection ratio plus the average coordination of a state in
phase space. An example of how to apply the method has been presented for a
very simple model that consists of two disks that moves in one dimension. A
similar procedure applied to a system of hard-disks shows a clear power law
scaling of the ratio between the boundary and the volume of the landscape. A
cross-over in the scaling exponents has been observed for a given packing
fraction. Near the freezing point, the boundary of the landscape scales as
the volume in phase space, and as a result the system tend to stay in
pockets of the phase-space. We speculate that the cross-over observed in the
scaling is related with the different kinds of relaxation processes of the
fluid. In future works, we will further explore this idea.

\textbf{Acknowledgments.} This work was supported by DGAPA UNAM project
IN108502, and National Science Foundation-CONACyT joint project 41538.

\end{document}